\documentclass[twocolumn,showpacs,preprintnumbers,amsmath,amssymb]{revtex4}

\usepackage{graphicx}
\usepackage{dcolumn}
\usepackage{bm}
\usepackage{amsmath}
\usepackage{url}

\begin{document}


\title{Density Fluctuation Effects on Collective Neutrino Oscillations\\in O-Ne-Mg Core-Collapse Supernovae}

\author{John F. Cherry$^{1}$}
\author{Meng-Ru Wu$^{2}$}
\author{J. Carlson$^{3}$}
\author{Huaiyu Duan$^{4}$}
\author{George M. Fuller$^{1}$}
\author{Yong-Zhong Qian$^{2}$}

\affiliation{$^{1}$Department of Physics, University of California, San Diego, La Jolla, California 92093, USA}
\affiliation{$^{2}$School of Physics and Astronomy, University of Minnesota, Minneapolis, MN 55455, USA}
\affiliation{$^{3}$Theoretical Division, Los Alamos National Laboratory, Los Alamos, New Mexico 87545, USA}
\affiliation{$^{4}$Department of Physics and Astronomy, University of New Mexico, Albuquerque, New Mexico 87131, USA}

\date{\today}

\begin{abstract}
We investigate the effect of matter density fluctuations on supernova collective neutrino flavor oscillations.  In particular, we use full multi-angle, 3-flavor, self-consistent simulations of the evolution of the neutrino flavor field in the envelope of an O-Ne-Mg core collapse supernova at shock break-out (neutrino neutronization burst) to study the effect of the matter density \lq\lq bump\rq\rq\ left by the He-burning shell.  We find a seemingly counterintuitive increase in the overall $\nu_{\rm e}$ survival probability created by this matter density feature.  We discuss this behavior in terms of the interplay between the matter density profile and neutrino collective effects.  While our results give new insights into this interplay, they also suggest an immediate consequence for supernova neutrino burst detection: it will be difficult to use a burst signal to extract information on fossil burning shells or other fluctuations of this scale in the matter density profile.  Consistent with previous studies, our results also show that the interplay of neutrino self-coupling and matter fluctuation could cause a significant increase in the $\nu_{\rm e}$ survival probability at very low energy.

\end{abstract}


\maketitle

\section{Introduction}
Core collapse supernovae are fantastic engines for the creation of large neutrino fluxes.  In turn, these fluxes can engender large scale, collective neutrino flavor oscillations deep in the supernova envelope (see Refs.~\cite{Fuller87,Notzold:1988fv,Pantaleone92,Fuller:1992eu,Qian93,Samuel:1993sf,Qian95,Kostelecky:1995rz,Samuel:1996rm,Pastor02,Pastor:2002zl,Sawyer:2005yg,Fogli07,Duan08,Dasgupta:2008qy}  and see the review in Ref.~\cite{Duan:2010fr} and references therein).  In this paper we investigate a puzzling aspect of collective neutrino flavor transformation in supernovae: in some cases a matter density fluctuation can ${\it increase}$ neutrino flavor transformation rather than decrease it as simplistic models including neutrino self-coupling seemingly predict.

Recent numerical simulations of neutrino flavor transformation in O-Ne-Mg core collapse supernovae~\cite{Cherry:2010lr} illustrate this conundrum.  The results of these calculations agreed phenomenologically with the standard theoretical frame work.  However, the final neutrino flavor distribution in these simulations revealed that in the normal neutrino mass hierarchy, neutrinos initially in mass state 3 (the heaviest mass eigenstate) hopped to lower mass eigenstates with much lower probability than would be predicted by simple theoretical models~\cite{Duan08,Dasgupta:2008qy}.

In this paper we analyze the results of multi-angle, self-consistent, 3-flavor simulations of neutrino flavor evolution in the neutronization neutrino burst of an O-Ne-Mg supernova.  Here we study the particular case of flavor evolution of a pulse of primarily electron flavor neutrinos with an average energy of $11\,\rm MeV$ and a peak luminosity of $1\times 10^{53}\,{\rm erg}\,{\rm s}^{-1}$ for a specified set of neutrino mixing parameters and emission spectra.  

To study why theoretical predictions of the neutrino mass state hopping rate differ from what is observed in our simulations we chose to vary only a single parameter in our model of the O-Ne-Mg supernova, the matter density profile.  This affords us an opportunity to conduct an interesting side investigation.  We explore the possibility that the neutrino signal from this model could be used to detect features in the matter density profile of the supernova, assuming a knowledge of neutrino mixing parameters.

Terrestrial experiments, like the proposed long baseline neutrino experiments, hold great promise for revealing key neutrino flavor mixing parameters, such as the value of $\theta_{13}$ and the neutrino mass hierarchy.  These as yet unmeasured quantities influence how neutrinos change their flavors in the core collapse supernova environment.

If experiments can reveal neutrino mixing parameters, it stands to reason that the signal from a supernova could be used as a probe of supernova physics.  There is a rich physical interplay between the hydrodynamic motion and nuclear abundances in a supernova and the neutrino flux streaming out from the proto-neutron star at it's heart.  Armed with a refined understanding of neutrino flavor transformation physics, it is reasonable to ask whether the supernova neutrino signal could be used as a probe of the matter density profile in a supernova at times and depths that are impossible to measure with optical observations.

For the purposes of this study we have chosen the following neutrino mixing parameters: neutrino mass squared differences $\Delta m^{2}_{\odot} = 7.6\times10^{-5}\, \rm eV^{2}$, $\Delta m^{2}_{\rm atm} = 2.4\times10^{-3}\, \rm eV^{2}$; vacuum mixing angles $\theta_{12} = 0.59$, $\theta_{23} = \pi/4$, $\theta_{13} = 0.1$; and CP-violating phase $\delta = 0$.  Here we will concentrate on the normal neutrino mass hierarchy.   Along with this we model the neutronization neutrino pulse to be a pure electron neutrino Fermi-Dirac spectrum with average energy $\langle E_{\nu}\rangle =11\,\rm MeV$, a degeneracy parameter $\eta = 3$, and luminosity $L = 1.0\times 10^{53}\,{\rm erg}\,{\rm s}^{-1}$.

Previous simulations~\cite{Cherry:2010lr} and semi-analytic work~\cite{Duan08,Dasgupta:2008qy} agree broadly on the theoretical framework that should describe the flavor evolution of neutrinos in this case.  In section $\rm II$ we discuss neutrino flavor transformation with and without a matter density fluctuation.  In section $\rm III$ we discuss the methodology of our numerical calculations, while in section $\rm IV$ we outline a theoretical framework for collective neutrino oscillations in this regime.  We give an analysis of our numerical results in section $\rm V$ and conclusions in section $\rm VI$. 

\section{Neutrino Flavor Transformation With and Without Matter Fluctuations: A Case Study}

For our particular model, neutrinos emerging from the neutrinosphere initially are in pure electron flavor states.  As these neutrinos stream outward through the envelope, a collective effect known as the \lq\lq Neutrino Enhanced MSW\rq\rq\ effect (not to be confused with the MSW, Mikheyev-Smirnov-Wolfenstein effect) can produce mass state hopping of neutrinos out of the heaviest mass eigenstate (for both neutrino mass hierarchies) when matter densities are large.  Nominally, the hopping rate is set by a comparison of the scale height of the matter density  and a characteristic neutrino oscillation length in the resonance region.  When matter densities fall further, a second collective effect called the \lq\lq Regular Precession Mode\rq\rq\ begins.  All neutrinos in this mode begin to rotate around an effective field in flavor space at the same frequency, regardless of their energy.  Because of the $\nu_{\rm e}$ only emission of the neutronization neutrino pulse, this process conserves the total number of neutrinos occupying each mass eigenstate and produces the distinctive \lq\lq Flavor Swaps\rq\rq\ or \lq\lq Spectral Swaps\rq\rq\ seen in the final neutrino spectra.  

This last point is extremely important.  By conserving the number of neutrinos in each mass state, the flavor swaps freeze the flavor evolution history of the neutrinos into the final spectrum with a signature that stands out dramatically to an observer here on Earth.  This suggests that such an observer might be able to simply measure the swap energies in a detected supernova neutrino signal and work backward to construct an in-situ measurement of the matter density profile at high densities. 

Because the neutronization neutrino pulse of an O-Ne-Mg supernova has been well studied and is a relatively simple case of flavor swap formation, it serves as a good test case to study our ability to extract information about the supernova envelope from a detected neutrino burst signal.  For the inverted neutrino mass hierarchy, the sequence of events of neutrino flavor transformation produce only a single swap because only one mass state level crossing is present~\cite{Duan08,Dasgupta:2008qy,Cherry:2010lr}.  For the normal neutrino mass hierarchy, two swaps are produced because two separate level crossing populate all three mass eigenstates with neutrinos (for small $\theta_{13}$ this can be reduced to a single swap via the complete depopulation of mass state 3)~\cite{Duan08,Dasgupta:2008qy,Cherry:2010lr}. 

The density profile used for this study comes from a set of simulations by Nomoto~\cite{Nomoto84,Nomoto87}.  This profile is typical of what a mid-collapse O-Ne-Mg supernova might produce at the epoch of the neutronization neutrino pulse, $\sim 10\,\rm ms$ post bounce.  A feature that this profile possesses is a small bump in the matter density which is created by the star's helium burning shell in a range of radius bounded by $r \backsimeq 1080\, {\rm km} - 1100\,{\rm km}$ .  This feature is known to cause neutrinos at low energies to pass through multiple MSW resonances at the $\Delta m^{2}_{\rm atm}$ mass scale, and has been discussed in~\cite{Duan08,Dasgupta:2008qy,Cherry:2010lr}.  

\begin{figure}[!h]
\centering
\includegraphics[scale=.72]{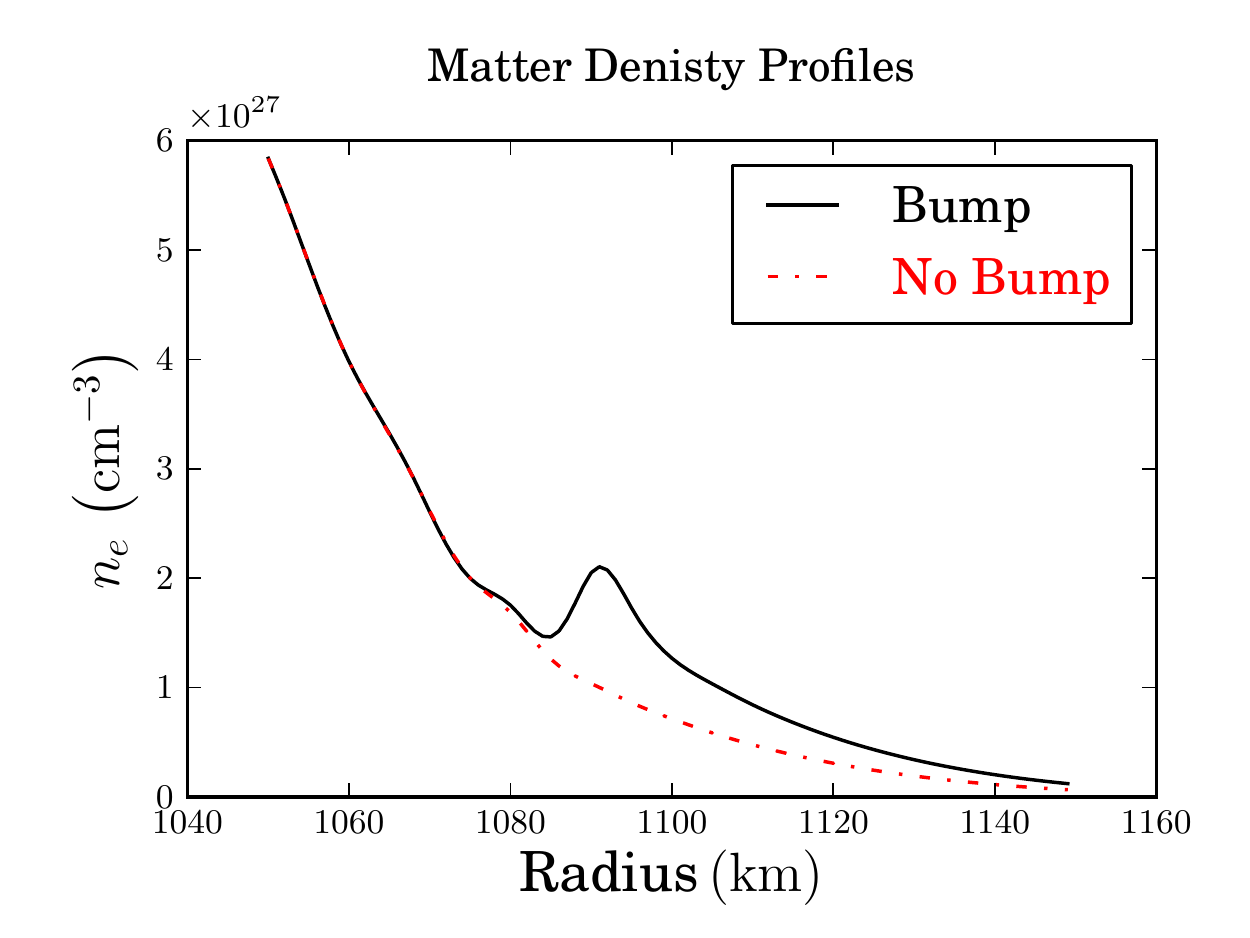}
\caption{The electron number density for two matter profiles plotted as functions of radius in the resonance region for the $\Delta m^{2}_{\rm atm}$ mass state splitting.  The solid line indicates the original matter density profile of Refs.~\cite{Nomoto84,Nomoto87}, called Bump.  The dashed-dotted line indicates the artificial density profile with the bump artificially removed, called No Bump.}
\label{fig:profile}
\end{figure}

\begin{figure}[!h]
\centering
\includegraphics[scale=.72]{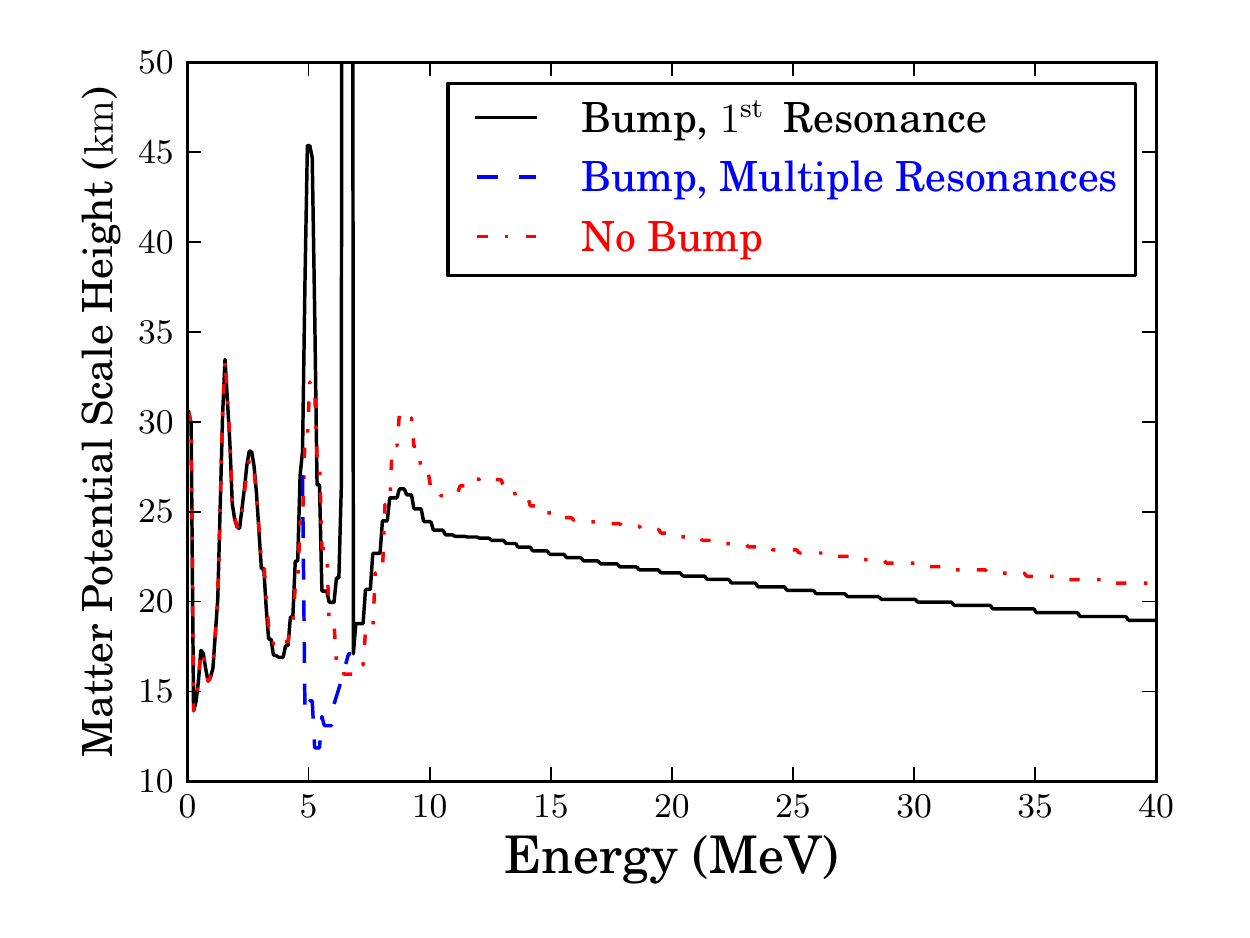}
\caption{The scale height of the neutrino-electron forward scattering potential evaluated at the MSW resonance location for each neutrino energy.  The solid line indicates the first MSW resonance scale height of neutrinos moving through the original matter density profile of Refs.~\cite{Nomoto84,Nomoto87}, while the dashed line indicates the scale heights of the multiple resonances.  The dash-dotted line indicates the MSW resonance scale heights of neutrinos moving through the No Bump density profile.}
\label{fig:SHMSW}
\end{figure}

To test our ability to detect a simple feature such as this, we conducted a pair of simulations.  The first with the original density profile, called \lq\lq Bump\rq\rq , and the second with a synthetic density profile where the He burning shell feature has been removed, called \lq\lq No Bump\rq\rq .  Figure~\ref{fig:profile} shows the electron number densities with these two profiles plotted side by side.  For the neutrino-electron forward scattering potential $H_{\rm e}$ (hereafter the \lq\lq matter\rq\rq\ potential) the associated scale height at resonance is, 
\begin{equation}
{\cal {H}} = \left| \frac{1}{H_{\rm e}}\frac{dH_{\rm e}}{dr}\right|^{-1}_{\rm res}.
\end{equation}
Figure~\ref{fig:SHMSW} shows the matter potential scale height for both profiles, evaluated at the MSW resonance location for each neutrino energy bin.  

\begin{figure*}
\centering
\includegraphics[scale=.90]{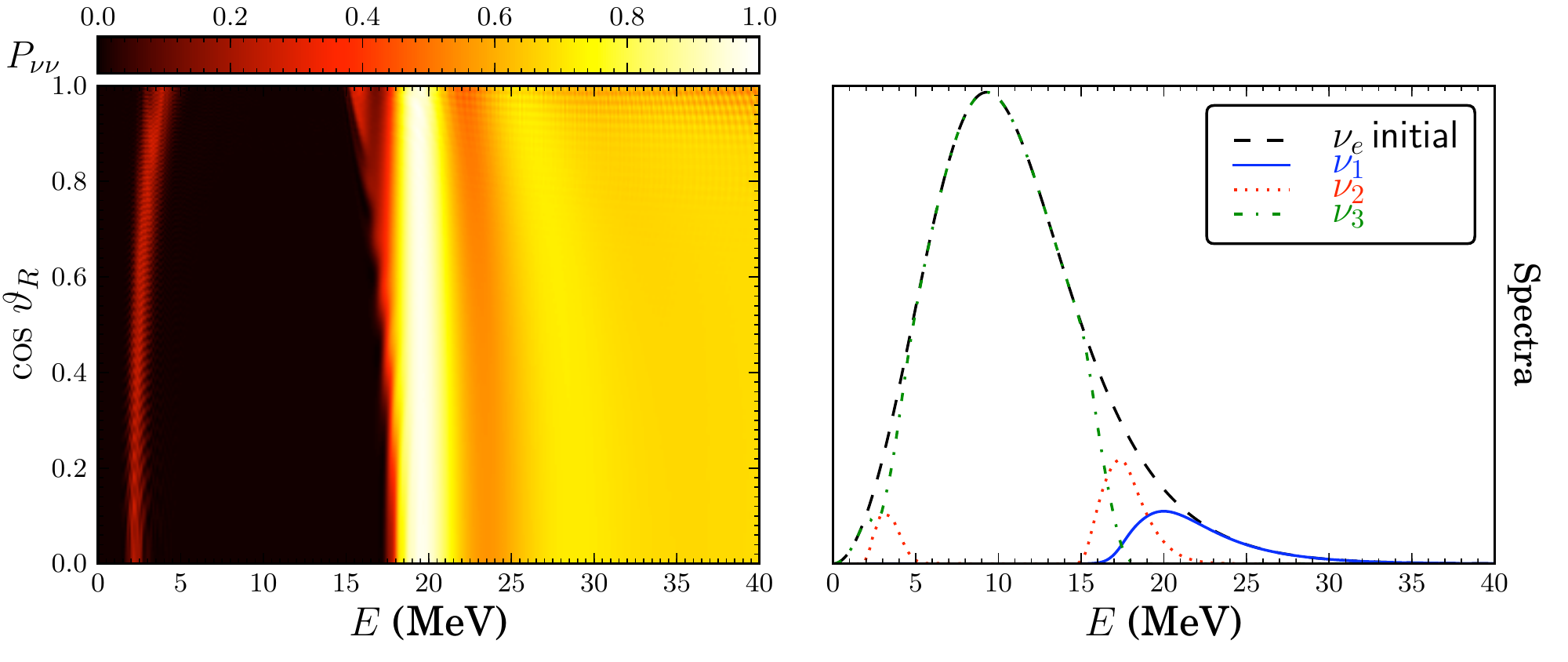}
\caption{Bump.  Left panel: electron neutrino survival probability $P_{\nu_{\rm e}\nu_{\rm e}}$ (color/shading key at top left) for the normal mass hierarchy is shown as a function of cosine of emission angle, $\cos{\vartheta_{\rm R}}$, and neutrino energy, $E$ in MeV, plotted at a radius of $r=5000\,{\rm km}$.  Right:  mass basis (key top right, inset) emission angle-averaged neutrino energy distribution functions versus neutrino energy, $E$.  The dashed curve gives the initial $\nu_{\rm e}$ emission angle-averaged energy spectrum.  A kink in the density profile used, taken from Refs.~\cite{Nomoto84,Nomoto87}, leads to multiple MSW resonances for low energy neutrinos.}
\label{fig:Pbump}
\end{figure*}

\begin{figure*}
\centering
\includegraphics[scale=.90]{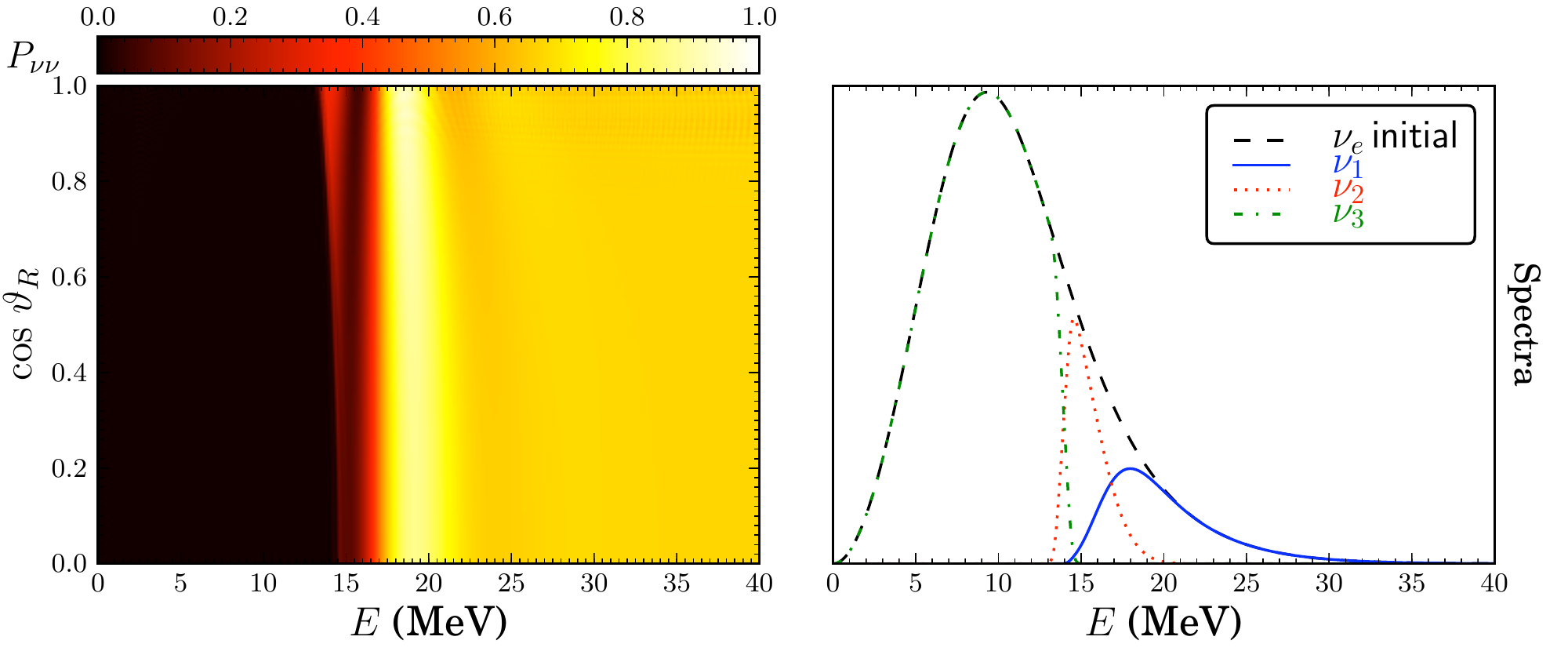}
\caption{No Bump.  Left panel: electron neutrino survival probability $P_{\nu_{\rm e}\nu_{\rm e}}$ (color/shading key at top left) for the normal mass hierarchy is shown as a function of cosine of emission angle, $\cos{\vartheta_{\rm R}}$, and neutrino energy, $E$ in MeV, plotted at a radius of $r=5000\,{\rm km}$.  Right:  mass basis (key top right, inset) emission angle-averaged neutrino energy distribution functions versus neutrino energy, $E$.  The dashed curve gives the initial $\nu_{\rm e}$ emission angle-averaged energy spectrum.  The kink in the density profile taken from Refs.~\cite{Nomoto84,Nomoto87}, has been artificially removed from the density profile used in this simulation.}
\label{fig:Pnobump}
\end{figure*}

Figures~\ref{fig:Pbump} and~\ref{fig:Pnobump} show the results of our calculations for the flavor transformation of electron neutrinos emitted during the neutronization neutrino burst.  Figure~\ref{fig:Pbump} shows the results of the original, Bump, density profile.  Figure~\ref{fig:Pnobump} shows the results of a simulation using the No Bump density profile. 

The aim of the second simulation, with the No Bump density profile, was to study whether we could detect a signature from features of the matter density profile using the neutrino flavor transformation signal.  Post processing of this data led to a surprise.  The total number of neutrinos that remain in the heavy mass eigenstate decreases when the bump in the density profile is removed.  Explicitly, the heavy mass eigenstate (mass state 3) survival probability, $P_{\rm H}$, for the two cases are $P^{\rm Bump}_{\rm H} = 0.852$ and $P^{\rm No\ \rm Bump}_{\rm H} = 0.759$.  This is a counterintuitive result.   We expected that the removal of the bump from the original density profile would have produced flavor evolution that was more adiabatic, leading to a greater survival probability for the No Bump profile.  

\begin{figure}[!h]
\centering
\includegraphics[scale=.72]{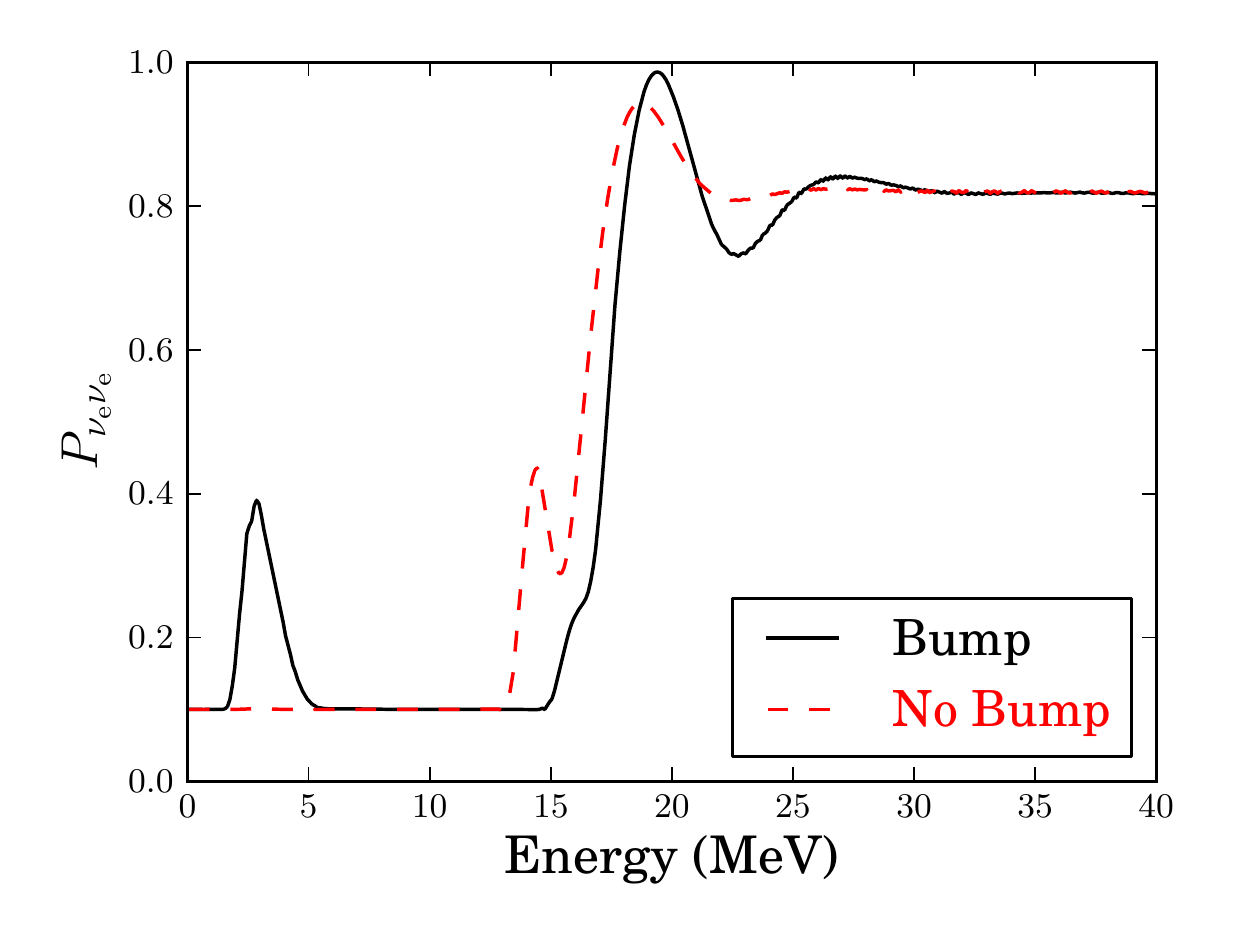}
\caption{The electron neutrino survival probability as a function of neutrino energy shown at the final radius $r=5000\, \rm km$.  The solid line indicates the survival probability for neutrinos after moving through the Bump profile, while the dashed line indicates the survival probability for neutrinos after moving through the No Bump profile.}
\label{fig:Peecomp}
\end{figure}

The effect that the removal of the helium burning shell has on the $\nu_{\rm e}$ survival probability is shown in Figure~\ref{fig:Peecomp}.  Note in this figure that there is an enhanced survival probability for $\nu_{\rm e}$'s at low energy for the Bump profile, and that the flavor swap energy is lowered for neutrinos in the No Bump profile.  

Collective flavor evolution prior to the onset of the regular precession mode was thought to be described by the Synchronous MSW effect, as opposed to the Neutrino Enhanced MSW effect~\cite{Duan08,Dasgupta:2008qy,Cherry:2010lr}.  While there is a technical difference between these two modes of neutrino flavor transformation, the neutrino flavor transformation survival probabilities (and consequently the Swap energies) are quite similar for both cases, which led to the initial confusion.  In both cases the resultant value of $P_{\rm H}$ is determined by the evolution of a single representative neutrino flavor state, which is the flavor state of the collective ensemble of neutrinos.  Ideally, neutrinos remain so closely aligned with this collective flavor state that they do not \lq\lq feel\rq\rq\ the neutrino self coupling potential and, as a result, they behave as a single neutrino experiencing the matter-driven MSW effect. 

\section{Methodology}
The simulation results analyzed here were produced with two numerical codes used for simulating neutrino flavor evolution.  These codes, the FLAT code and the BULB code, and related schemes to solve for the flavor evolution of core collapse supernova neutrinos, are discussed in Ref.~\cite{Duan:2008eb}. 

In order to parallelize the nonlinearly-coupled differential equations which describe neutrino flavor evolution, BULB employs a spherically symmetric representation of the region above the neutrino sphere.  All neutrinos are assumed to be emitted from a hard spherical shell, and propagate through a one dimensional distribution of matter.  These choices allow the neutrino emission to be broken down and grouped by species, energy, and emission angle at the neutrino sphere.  Here we define the emission angle to be the angle between the neutrino direction and the vector normal to the surface of the neutrino sphere at the neutrino emission point.  To initialize the simulation, neutrinos are allocated to each energy-angle bin according to the species-specific luminosity and neutrino energy spectrum characteristics.  From there, BULB employs a second order predictive-corrective algorithm to compute the flavor evolution of the neutrino states.  In order to check for convergence, a second round of computations are made with a step size ${\Delta t}/{2}$.  The final neutrino flavor states from the double iteration are then compared to the final states of the original step to verify that they agree to within a predefined error tolerance, usually chosen to be 1 part in $10^{8}$.  Convergence of the overall calculation is checked by comparing results with different error tolerances and differing numbers of energy and angle bins.

We have validated our simulations by performing them with both codes and comparing our results~\cite{Duan:2008eb,Cherry:2010lr}.  Using the same set of initial conditions, both codes agree with each other at the level of $0.1\%$ when comparing the final neutrino flavor states.  For a more detailed description of the inner workings of the BULB and FLAT codes, see Refs.~\cite{Duan06a,Duan:2008eb,Cherry:2010lr}.

\section{Theory}

For our simulations we assume a pure $\nu_{\rm e}$ burst emitted from the neutrino sphere ar $R_{\nu}=60\,{\rm km}$ with a total luminosity of $L_{\nu}=10^{53}\,{\rm erg}\,{\rm s}^{-1}$ and a normalized spectrum
\begin{equation}
f_{\nu}\left( E\right)=\frac{1}{F_{2}\left(\eta_{\nu}\right)T^{3}_{\nu}} \\
\frac{E^{2}}{{\rm exp}\left( E/T_{\nu} - \eta_{\nu} \right) + 1},
\end{equation}
where $\eta_{\nu} = 3$ and $T_{\nu} = 2.75\,{\rm MeV}$.  This corresponds to an average $\nu_{\rm e}$ energy $\langle E_{\nu} \rangle = F_{3}\left(\eta_{\nu}\right)T_{\nu}/F_{2}\left(\eta_{\nu}\right) = 11\,{\rm MeV}$.  Here
\begin{equation}
F_{n}\left(\eta_{\nu}\right) = \int_{0}^{\infty} \frac{x^{n}}{{\rm exp}\left(x - \eta_{\nu}\right) + 1}dx\ .
\end{equation}
In the single-angle approximation, the effective total neutrino number density at $r > R_{\nu}$ is
\begin{equation}
n_{\nu}\left(r\right) = \frac{L_{\nu}}{4\pi R^{2}_{\nu}\langle E_{\nu}\rangle}\left[ 1-\sqrt{1 \\ 
-\left( R_{\nu}/r\right)^{2}}\right]^{2} \approx \frac{L_{\nu}R_{\nu}^{2}}{16\pi\langle E_{\nu}\rangle r^{4}}\ ,
\end{equation}
where the approximate equality holds for $r\gg R_{\nu}$, and where we set $\hbar = c = 1$.

Because the salient features of our results are confined to the $\delta m^{2}_{\rm atm}$ mass squared mixing scale, we will confine the following discussion to a two neutrino flavor mixing scheme.  Following the convention of~\cite{Duan06c}, we take each neutrino flavor state with energy $E_{\nu}$ and represent it as a three dimensional neutrino flavor isospin (NFIS), 
\begin{equation}
\mathbf s_\omega = \left\{\begin{matrix}
\hat{\mathbf e}^{\rm f}_{\rm z}/2, & \rm for\ \nu_{\rm e}, \\
 -\hat{\mathbf e}^{\rm f}_{\rm z}/2, & \rm for\ \nu_{\rm x},\end{matrix}\right.
\end{equation}
where $\hat{\mathbf e}^{\rm f}_{\rm z}$ is the unit vector in the z-direction for the neutrino flavor basis and $\omega$ is the vacuum oscillation frequency $\omega = \delta m^{2}_{atm}/2E_{\nu}$.  We focus on the normal mass hierarchy and we take the effective vacuum mixing angle $\theta_{\rm v}\approx\theta_{13} = 0.1$.  The evolution of a NFIS $\mathbf s_{\omega}$ is governed by
\begin{multline}
\frac{d}{dr}\mathbf s_{\omega} = \mathbf s_{\omega} \times \\ 
\left[ \omega \mathbf H_{\rm v} 
+ \mathbf H_{\rm e} - \mu\left( r\right) \int_{0}^{\infty} \mathbf s_{\omega^\prime}f_{\nu}\left(
E_{\omega^\prime}\right)dE_{\omega^\prime}\right]\ ,
\label{FullHam}
\end{multline}
where $\mathbf H_{\rm v} = \cos{2\theta_{\rm v}}\hat{\mathbf e}^{\rm f}_{\rm z} - \sin{2\theta_{\rm v}}\hat{\mathbf e}^{\rm f}_{\rm x}$, $\mathbf H_{\rm e} = -\sqrt{2}G_{\rm F}n_{\rm e}\left( r\right)\hat{\mathbf e}^{\rm f}_{\rm z}$, $\mu\left( r\right) = 2\sqrt{2}G_{\rm F}n_{\nu}\left( r\right)$, and $E_{\omega^\prime} = \delta m^{2}_{\rm atm}/2\omega^\prime$.  For convenience, we define
\begin{equation}
g\left(\omega\right) \equiv \frac{\delta m^{2}_{\rm atm}}{2\omega}f_{\nu}\left(E_{\omega}\right)
\label{littleg}
\end{equation}
and
\begin{equation}
\mathbf S \equiv \int_{0}^{\infty} \mathbf s_{\omega}f_{\nu}\left(E_{\omega}\right)dE_{\omega} \\
= \int_{0}^{\infty} \mathbf s_{\omega}g\left(\omega\right)d\omega\ .
\label{Scol}
\end{equation}
It follows from Eqns.~\eqref{FullHam}-\eqref{Scol} that 
\begin{equation}
\frac{d}{dr}\mathbf S = \int_{0}^{\infty} \omega g\left(\omega\right)\mathbf s_{\omega} d\omega \times \\
\mathbf H_{\rm v} + \mathbf S \times \mathbf H_{\rm e}\ .
\end{equation}

As $g\left(\omega\right)$ is concentrated in a finite range of $\omega$, to zeroth order we approximate $g\left(\omega\right) \approx \delta\left(\omega - \langle\omega\rangle\right)$, where $\langle\omega\rangle = \int_{0}^{\infty} \omega g\left(\omega\right)d\omega$.  This is a fair approximation for the particular case we treat, namely, low energy $\nu_{\rm e}$'s in the neutronization burst ($\langle E_{\nu}\rangle= 11\, \rm MeV$).  With this approximation, the zeroth-order mean field $\mathbf S^{(0)}$ is defined through
\begin{equation}
\frac{d}{dr}\mathbf S^{(0)} = \mathbf S^{(0)} \times \left[\langle\omega\rangle\mathbf H_{\rm v} + \\
\mathbf H_{\rm e}\right] \equiv \mathbf S^{(0)} \times \mathbf H_{\rm MSW}\ .
\label{S_zero_EOM}
\end{equation}

The evolution of $\mathbf S^{(0)}$ is the behavior of the system in the high luminosity, \lq\lq Synchronized\rq\rq , limit.  (In the Synchronous MSW effect, all $\mathbf s_{\omega}$ are aligned with $\mathbf S$, and orbit around it.  Note, this idealized situation does not occur in the presence of matter~\cite{Duan07a}.)  For the neutronization neutrino burst luminosity and matter density profile we use, our calculations take place below this luminosity regime.  We do not observe individual $\mathbf s_{\omega}$ orbiting $\mathbf S$~\cite{Cherry:web}, but we do observe that individual $\mathbf s_{\omega}$ remain closely aligned to $\mathbf S$ and $\mathbf S^{(0)}$.

The evolution of $\mathbf S^{(0)}$ is the same as that of an idealized $\nu_{\rm e}$ with $E_{\nu} = \delta m^{2}_{\rm atm}/2\langle\omega\rangle = 8.53\, {\rm MeV}$ undergoing the usual MSW effect.  This is the collective NFIS that all the the neutrinos will follow during the Neutrino Enhanced MSW effect.  Now we can approximately solve for the evolution of $\mathbf s_{\omega}$ from 
\begin{equation}
\frac{d}{dr}\mathbf s_{\omega} \approx \mathbf s_{\omega} \times \left[\omega\mathbf H_{\rm v} \\
 + \mathbf H_{\rm e} - \mu\left( r\right)\mathbf S^{(0)}\right]\ .
\end{equation}
We can use the solution of Equation~\ref{S_zero_EOM} to obtain the first-order mean field $\mathbf S^{(1)}$ from the definition of $\mathbf S$ in the first expression in Equation~\ref{Scol}, and then re-calculate the evolution of $\mathbf s_{\omega}$ from the above Equation~\ref{S_zero_EOM} but with $\mathbf S^{(1)}$ replacing $\mathbf S^{(0)}$.  This procedure can be repeated until the results converge.

\begin{figure*}[!htp]
\centering
\includegraphics[scale=.90]{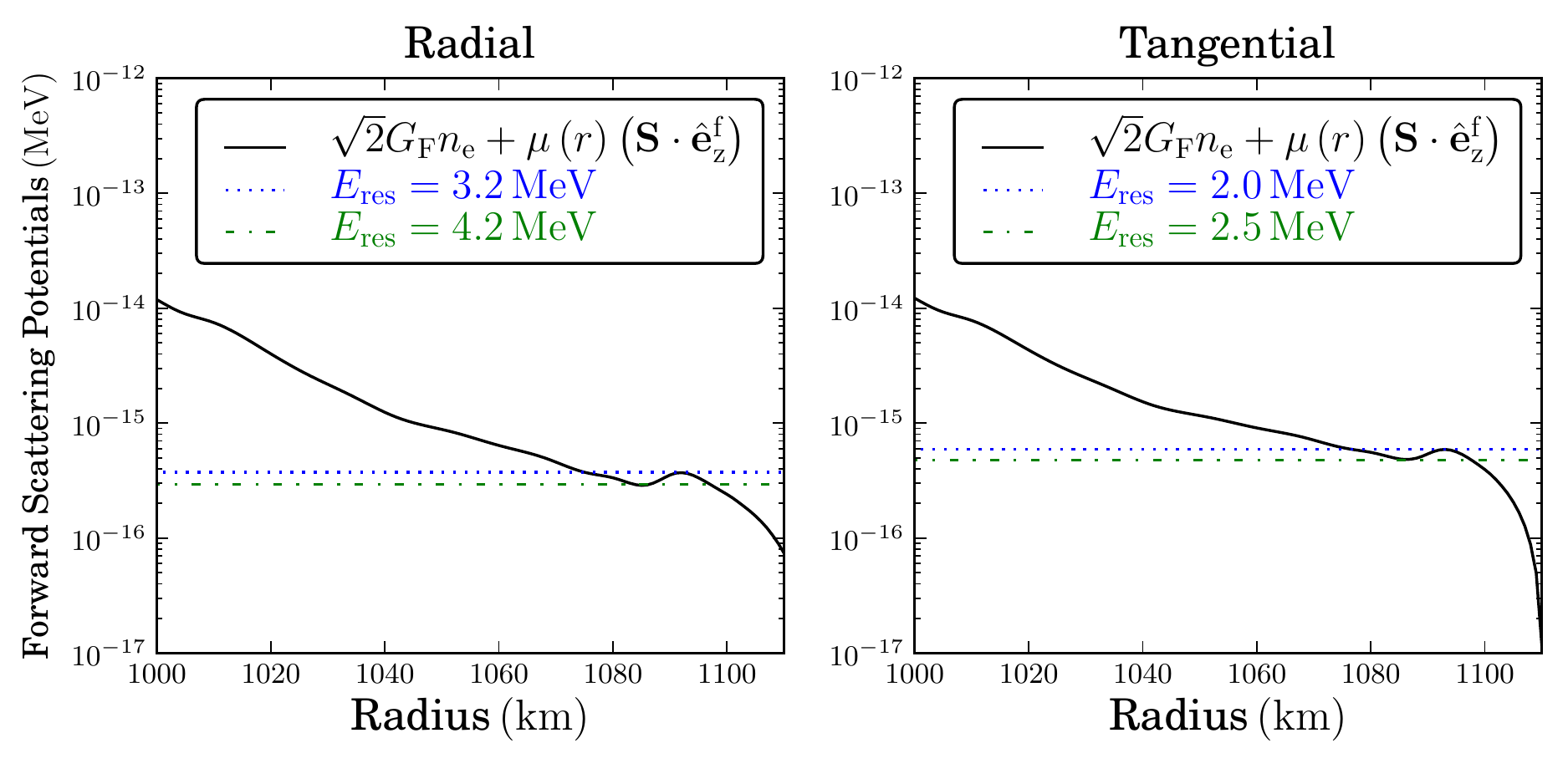}
\caption{Left panel:  the combined matter and neutrino self-coupling forward scattering potentials experienced by radially emitted neutrinos as a function of radius.  Right:  the combined matter and neutrino self-coupling forward scattering potentials experienced by tangentially emitted neutrinos as a function of radius.  The dotted and dashed-dotted lines indicate the upper and lower neutrino energies, respectively, to experience multiple neutrino-background enhanced MSW resonances in the desynchronized limit.}
\label{fig:RESWINDOW}
\end{figure*}

Of course, the procedure outlined above is not recommended as a numerical method, but instead points to an analytic approach to understand collective oscillations.  Based on the MSW effect, $\mathbf S^{(0)}$ goes through resonance at $n_{\rm e}\approx 1.09\times 10^{27}\,{\rm cm}^{-3}$ corresponding to $r\approx 1100\, {\rm km}$ in both simulations.  For simplicity, let us consider a smooth distribution for $n_{\rm e}\left( r\right)$.  Before the resonance, the evolution of $\mathbf S^{(0)}$ is somewhat adiabatic and we can take
\begin{align}
\mathbf S^{(0)} & \approx -\frac{\mathbf H_{\rm MSW}}{2|\mathbf H_{\rm MSW}|} \\
 & \approx -\frac{1}{2}\left(\cos{2\theta_{\rm m}}\hat{\mathbf e}^{\rm f}_{\rm z} - 
\sin{2\theta_{\rm m}}\hat{\mathbf e}^{\rm f}_{\rm x}\right)\ ,
\end{align}
where
\begin{multline}
\cos{2\theta_{\rm m}} = \\ 
\frac{\langle\omega\rangle\cos{2\theta_{\rm v}} - \sqrt{2}G_{\rm F}n_{\rm e}}{\sqrt{\left(\langle\omega\rangle\cos{2\theta_{\rm v}} - \sqrt{2}G_{\rm F}n_{\rm e}\right)^{2} + 
\left(\langle\omega\rangle\sin{2\theta_{\rm v}}\right)^{2}}}\ ,
\end{multline}
\begin{multline}
\sin{2\theta_{\rm m}} = \\
 \frac{\langle\omega\rangle\sin{2\theta_{\rm v}}}{\sqrt{\left(\langle\omega\rangle\cos{2\theta_{\rm v}} - \sqrt{2}G_{\rm F}n_{\rm e}\right)^{2} + 
\left(\langle\omega\rangle\sin{2\theta_{\rm v}}\right)^{2}}}\ .
\end{multline}
The evolution of $\mathbf s_{\omega}$ at densities higher than the resonance density for $\mathbf S^{(0)}$ is then governed by
\begin{widetext}
\begin{align}
\frac{d}{dr} \mathbf s_{\omega} & \approx \mathbf s_\omega \times \left[\left(\omega\cos{2\theta_{\rm v}} - \sqrt{2} G_{\rm F}n_{\rm e} +\frac{\mu}{2}\cos{2\theta_{\rm m}}\right)\hat{\mathbf e}^{\rm f}_{\rm z} - 
\left(\omega\sin{2\theta_{\rm v}} + \frac{\mu}{2}\sin{2\theta_{\rm m}}\right)\hat{\mathbf e}^{\rm f}_{\rm x}\right] \\
 & \equiv \mathbf s_\omega \times \mathbf H_\omega\ .
\end{align}
\end{widetext}
This equation also defines $\mathbf H_{\omega}$.  Note that the NFIS evolution described by this equation is similar to the usual MSW effect but with modified diagonal and off-diagonal terms.  Note especially that the off-diagonal term $\left(\mu /2\right)\sin{2\theta_{\rm m}} = \mu /2$ is large at the resonance location for $\mathbf S^{(0)}$.  If $\mathbf s_\omega$ evolves adiabatically, then
\begin{align}
\mathbf s_\omega & \approx -\frac{\mathbf H_\omega}{2|\mathbf H_\omega |} \\
 & \approx -\frac{1}{2}\left(\cos{2\theta_{\omega}}\hat{\mathbf e}^{\rm f}_{\rm z} -
\sin{2\theta_{\omega}}\hat{\mathbf e}^{\rm f}_{\rm x}\right)\ ,
\end{align}
where
\begin{widetext}
\begin{equation}
\label{costhw}
\cos{2\theta_\omega} = \frac{\langle\omega\rangle\cos{2\theta_{\rm v}} - \sqrt{2}G_{\rm F}n_{\rm e} + \left(\mu /2\right)\cos{2\theta_{\rm m}}}{\sqrt{\left(\langle\omega\rangle\cos{2\theta_{\rm v}} -\\
 \sqrt{2}G_{\rm F}n_{\rm e}+ \left(\mu /2\right)\cos{2\theta_{\rm m}}\right)^{2} + \\
\left(\langle\omega\rangle\sin{2\theta_{\rm v}} + \left(\mu /2\right)\sin{2\theta_{\rm m}}\right)^{2}}}\ ,
\end{equation}
\begin{equation}
\sin{2\theta_\omega} = \frac{\langle\omega\rangle\sin{2\theta_{\rm v}} + \left(\mu /2\right)\sin{2\theta_{\rm m}}}{\sqrt{\left(\langle\omega\rangle\cos{2\theta_{\rm v}} -\\
 \sqrt{2}G_{\rm F}n_{\rm e}+ \left(\mu /2\right)\cos{2\theta_{\rm m}}\right)^{2} + \\
\left(\langle\omega\rangle\sin{2\theta_{\rm v}} + \left(\mu /2\right)\sin{2\theta_{\rm m}}\right)^{2}}}\ .
\end{equation}
\end{widetext}
If $\mathbf s_\omega$ goes through resonance non-adiabatically at densities above the resonance density for $\mathbf S^{(0)}$, then subsequently $\mathbf s_\omega$ will no longer stay anti-aligned with $\mathbf H_\omega$.  This change in alignment corresponds to neutrinos jumping between mass states.  Confirmation of this simple picture is borne out by our numerical simulations as can be seen in Figures~\ref{fig:Pbump} and~\ref{fig:Pnobump}, where the Bump profile exhibits a population of low energy mass state 2 ($\nu_{2}$) neutrinos.  In the Bump density profile, Figure~\ref{fig:Pbump}, the helium burning shell produces multiple MSW-like resonances for low energy neutrinos.  These multiple resonances,illustrated in Figure~\ref{fig:RESWINDOW}, are non-adiabatic, engendering further loss of alignment.   This leads to a population of low energy neutrinos occupying mass state 2, which do not recover their alignment with $\mathbf H_\omega$.  

By way of contrast, in Figure~\ref{fig:Pnobump} there is no population of low energy neutrinos in mass state 2 because the absence of the helium burning shell in the No Bump profile makes the evolution of these neutrinos adiabatic, hence they remain aligned with $\mathbf H_\omega$.  In both cases, NIFS's $\mathbf s_\omega$ over a wide range of $\omega$ experience significant evolution at densities higher than the resonance density of $\mathbf S^{(0)}$.

If $\mathbf S^{(0)}$ goes through resonance adiabatically, the above description of NFIS evolution can be extended to lower densities.  Note that $\cos{2\theta_{\rm m}}$ changes from $\approx -1$ at high density to $0$ at resonance and to $\approx \cos{2\theta_{\rm v}}$ at low density, consistent with a simple MSW picture.  The diagonal term$\sqrt{2}G_{\rm F}n_{\rm e} - \left(\mu /2\right)\cos{2\theta_{\rm m}}$ becomes $0$ at some radius and all $\mathbf s_\omega$ go through resonance before this radius.

However, if $\mathbf S^{(0)}$ goes through resonance non-adiabatically, the situation becomes more complicated.  For illustration consider the regime of low $n_{\rm e}$, lower than the $\mathbf S^{(0)}$ resonance density.  The non-adiabatic evolution of $\mathbf S^{(0)}$ means that it no longer stays anti-aligned with $\mathbf H_{\rm MSW} \approx \langle\omega\rangle\mathbf H_{\rm v}$ in this regime.  Instead, we have
\begin{equation}
\mathbf S^{(0)} \cdot \hat{\mathbf H}_{\rm v} = \frac{1}{2}\cos{\alpha}  = P_{\rm hop} - \frac{1}{2}\ ,
\end{equation}
where $\alpha$ is the angle between $\mathbf S^{(0)}$ and the external field $\mathbf H_{\rm v}$, and $P_{\rm hop}$ is the probability for $\mathbf S^{(0)}$ to hop from being anti-aligned before resonance to being aligned with $\mathbf H_{\rm MSW}$ after resonance.  We can take
\begin{equation}
\mathbf S^{(0)} \approx \left(P_{\rm hop} -\frac{1}{2}\right)\hat{\mathbf H}_{\rm v} + \mathbf S^{(0)}_{\perp}\ ,
\end{equation}
where $\mathbf S^{(0)}_{\perp}$ is the component perpendicular to $\mathbf H_{\rm v}$ with a magnitude $\approx \sqrt{\left( 1/2\right)^2 -\left[ P_{\rm hop} - \left( 1/2\right)\right]^2}$.  Note that $\mathbf S^{(0)}_\perp$ precesses around $\mathbf H_{\rm v}$ with an angular velocity $-\langle\omega\rangle\mathbf H_{\rm v}$.  

In a frame co-precessing with $\mathbf S^{(0)}$, the evolution of $\mathbf s_\omega$ is governed by
\begin{widetext}
\begin{align}
\frac{d}{dr}\mathbf s_\omega & \approx \mathbf s_\omega \times \left[\left(\omega - \langle\omega\rangle\right)\mathbf H_{\rm v} - \mu\left( r\right)\mathbf S^{(0)}\right] \\
&\approx \mathbf s_\omega \times \{\left[\omega-\langle\omega\rangle-\mu\left( r\right)\left(P_{\rm hop} - \frac{1}{2}\right)\right]\mathbf H_{\rm v} - \mu\left( r\right)\mathbf S^{(0)}_\perp\} \\
&\equiv \mathbf s_\omega \times \mathbf H^{\rm co-pre}_\omega\ .
\end{align}
\end{widetext}
The evolution of $\mathbf s_\omega$ is expected to be adiabatic and, as a result, the angle between $\mathbf s_\omega$ and $\mathbf H^{\rm co-pre}_\omega$ stays fixed.  This angle depends on the relative directions of $\mathbf s_\omega$ and $\mathbf H^{\rm co-pre}_\omega$ right after the resonance of $\mathbf S^{(0)}$.  The latter direction depends on the exact direction of $\mathbf S^{(0)}_\perp$ coming out of the resonance.  At large radii, $\mathbf H^{\rm co-pre}_\omega$ simply becomes $\left(\omega - \langle\omega\rangle\right)\mathbf H_{\rm v}$.  Those $\mathbf s_\omega$ with $\omega < \langle\omega\rangle$ that are approximately aligned with $\mathbf H^{\rm co-pre}_\omega$ right after the resonance of $\mathbf S^{(0)}$ are nearly fully converted into $\nu_{\rm x}$.

Neutrinos participating in the Neutrino Enhanced MSW effect have NIFS's which are closely aligned with the collective field $\mathbf S$, so their flavor evolution will be a close match to that of $\mathbf S^{(0)}$~\cite{Duan07a}.  Specifically, they will co-precess with the effective field, even when the flavor evolution of $\mathbf S^{(0)}$ and $\mathbf S$ are non-adiabatic.  Examples of this co-precession can be viewed in  movies which are on our website~\cite{Cherry:web}.  

In this co-precession picture the probability to remain in the heavy mass eigenstate, $P_{\rm H}$, should depend on the three quantities relevant to the MSW evolution of $\mathbf S^{(0)}$: vacuum mixing angle  $\theta_{\rm v}$; collective oscillation frequency $\langle \omega \rangle$; and the matter potential scale height $R_{\rm H}$ at the location where $\mathbf S^{(0)}$ is at resonance, $\langle \omega_{H} \rangle\cos{2\theta_{\rm v}} = 2\sqrt{2}G_{\rm F}n_{\rm e}\left( r\right)$.  The first and second quantities are identical for both of these simulations.  Only  $R_{\rm H}$ changes when the burning shell feature is removed, with $R^{\rm Bump}_{\rm H} = 25.3\,{\rm km}$ and $R^{\rm No\ Bump}_{\rm H} = 28.5\,{\rm km}$.  

Following the evolution of $\mathbf S^{(0)}$ through the envelope of matter around the proto-neutron star, the probability of a neutrino with $\omega = \langle \omega \rangle$ to hop out of the heavy mass eigenstate will be given by the double exponential Landau-Zener hopping probability, $P_{\rm hop} = \left(1-P_{\rm H}\right)$, with:
\begin{equation}
P_{\rm hop} = \frac{\exp\left( 2\pi R_{\rm H} \langle \omega\rangle \cos^{2}\theta_{\rm v}\right) \\
 - 1}{\exp\left( 2\pi R_{\rm H} \langle \omega \rangle\right) - 1}.
\label{Phop}
\end{equation} 
Given that the critical scale height of the matter profile is slightly smaller for the Bump profile, the above equation implies that there should be a larger probability to hop out of mass state 3 in the presence of the helium burning shell.  This is why the total number of neutrinos remaining in mass state 3 naively is expected to increase for the No Bump simulation.  Equation~\ref{Phop} yields a prediction that for the Bump density profile $P^{\rm Predicted}_{\rm H} = 0.68$, and for the No Bump density profile $P^{\rm Predicted}_{\rm H} = 0.71$.

Using the final emission spectra from Figures~\ref{fig:Pbump} and~\ref{fig:Pnobump} to calculate $P_{\rm H}$ for both simulations, as we would in analyzing an actual supernova neutrino signal, produces somewhat different results.  For the No Bump calculation, $P^{\rm Observed}_{\rm H} = 0.76$.  This indicates a slightly more adiabatic than we had calculated above, and would lead an observer to deduce a larger matter scale height in the collapsing core of the supernova than was actually present.  More strikingly, the Bump profile exhibits $P^{\rm Observed}_{\rm H}=0.85$, implying fully $17\%$ more neutrinos remain in mass state 3 than predicted.  This would lead an observer interested in the envelope to grossly miss-calculate the electron density scale height, arriving at a number nearly twice the actual value.  We will endeavor to understand why it is that our model of neutrino flavor transformation seems to have led us astray when attempting to work backward from our observed signal to the matter density profile of the collapsing star.

\section{Analysis}

Though the simple neutrino transformation model presented above is successful in many respects, we do not attempt to provide an exhaustive proof of the model, only to point out that it offers a straightforward explanation of the puzzling aspects of our result.  Furthermore, our numerical calculations show that one must be careful in applying the theory of the Neutrino Enhanced MSW effect to infer information about the envelope of the proto-neutron star.

First, even in early (e.g. Ref.~\cite{Qian95}) treatments of neutrino flavor evolution it was evident that flavor diagonal neutrino-neutrino forward scattering potential, $B\left( r\right)$, would alter the position of the MSW resonance position for a given neutrino energy $E_{\nu}$ and mass splitting $\Delta m^{2}$ because the resonance condition is
\begin{equation}
\frac{\Delta m^{2}}{2E}\cos{2\theta_{\rm v}} = A\left( r\right)+B\left( r\right)\ ,
\label{BDSres}
\end{equation}
where $A\left( r\right) = \sqrt{2}G_{\rm F}n_{\rm e}\left( r\right)$ is the matter potential at radius $r$.  Neutrino propagation through the MSW resonance at the shifted position in general will result in an altered survival probability because the scale height of the combined potential will be different.  However, this simplistic analysis is completely inadequate because it is the neutrino-neutrino flavor off-diagonal potential which in part determines adiabaticity~\cite{Qian95,Fuller06}.  This potential, in turn, is sensitive to the relative $x-y$ plane phase angles of the individual neutrino NFIS's,  necessitating a self-consistent collective oscillation treatment.

\begin{figure}[!htp]
\centering
\includegraphics[scale=.72]{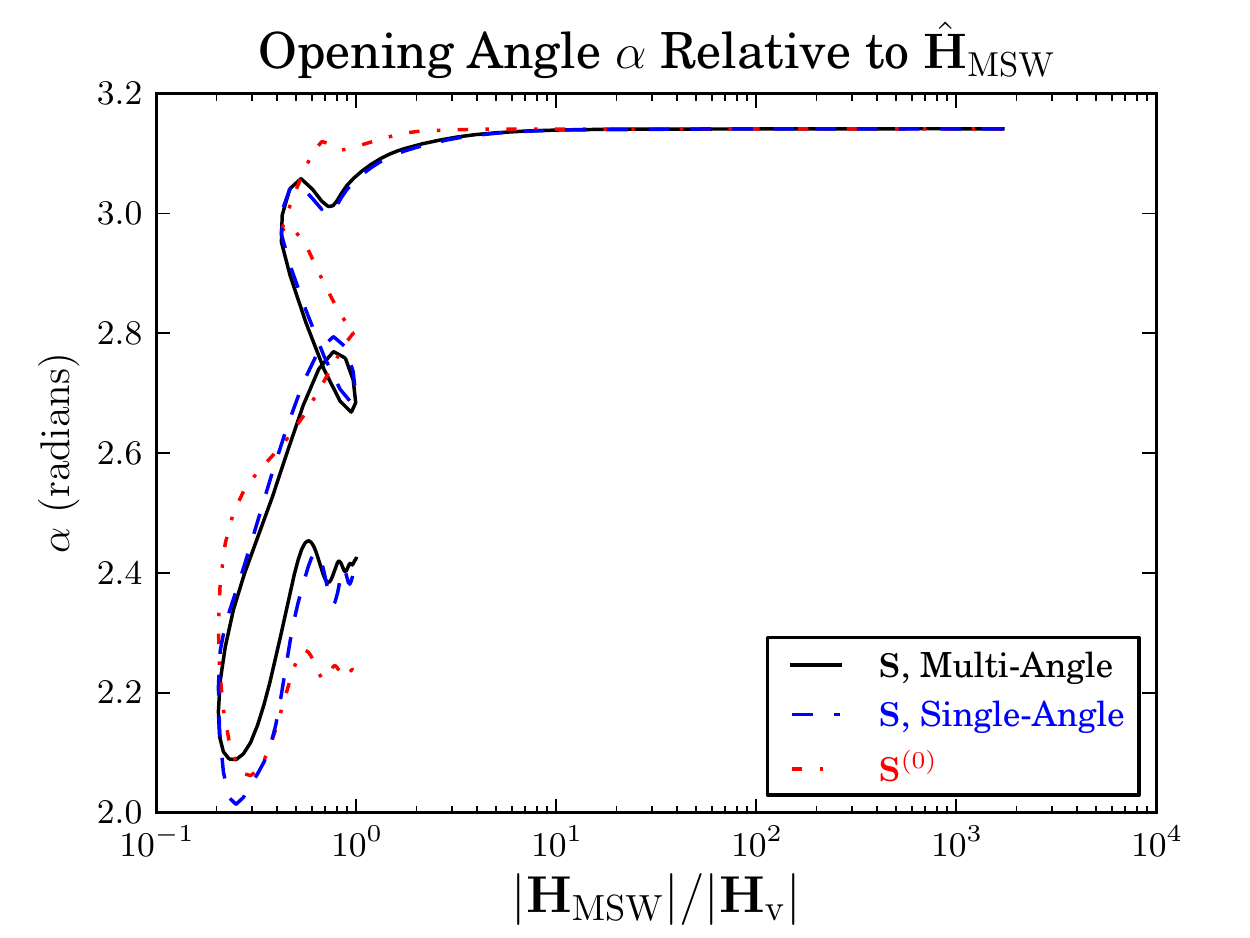}
\caption{Bump:  The opening angle $\alpha$ between the collective NFIS and $\mathbf H_{\rm MSW}$ for the Bump profile, plotted as a function of $\left|\mathbf H_{\rm MSW}\right|/\left|{\mathbf H}_{\rm v}\right|$ as the system moves through resonance.  The idealized NFIS (dotted-dashed line) shows the evolution of $\mathbf S^{(0)}$ in the absence of any neutrino self-coupling.  The solid line and dashed line show the evolution of $\mathbf S$ as calculated in multi-angle and single-angle simulations respectively, including the neutrino self-coupling potentials.}
\label{fig:ALGNBUMP}
\end{figure}

\begin{figure}[!htp]
\centering
\includegraphics[scale=.720]{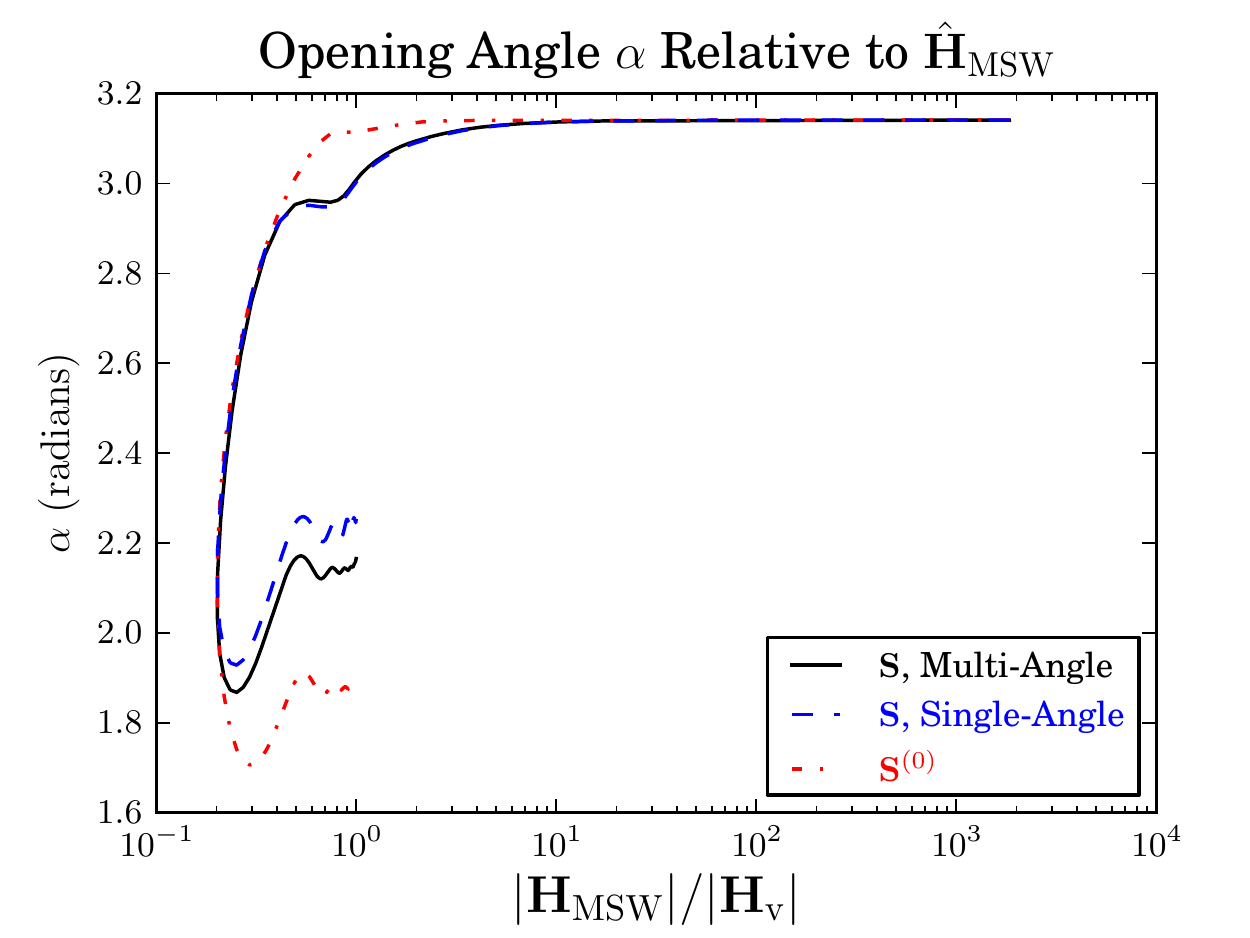}
\caption{No Bump:  The opening angle $\alpha$ between the collective NFIS and $\mathbf H_{\rm MSW}$ for the No Bump profile, plotted as a function of $\left|\mathbf H_{\rm MSW}\right|/\left|{\mathbf H}_{\rm v}\right|$  as the system moves through resonance.  The idealized NFIS (dotted-dashed line) shows the evolution of $\mathbf S^{(0)}$ in the absence of any neutrino self-coupling.  The solid line and dashed line show the evolution of $\mathbf S^{(0)}$ as calculated in multi-angle and single-angle simulations respectively, including the neutrino self-coupling potentials.}
\label{fig:ALGNNOBUMP}
\end{figure}

Shown in Figures~\ref{fig:ALGNBUMP} and~\ref{fig:ALGNNOBUMP} is the evolution of the Neutrino Enhanced MSW effect collective NFIS $\mathbf S$ and evolution of the zeroth order approximation $\mathbf S^{(0)}$.  Each of these are given for the Bump and No Bump profiles.  These figures show the opening angle between the collective NFIS and $\mathbf H_{\rm MSW}$.  To lowest order, the collective NFIS $\mathbf S$ from Eqn.~\ref{Scol} follows the alignment of $\mathbf S^{(0)}$, starting anti-aligned with $\mathbf H_{\rm MSW}$ and undergoing a mild change of alignment as the system passes through resonance, with the hopping probability given by $P_{\rm hop} = 1/2\left(1+\cos{\alpha}\right)$.  However, there is a small difference between the final alignment of $\mathbf S^{(0)}$ and $\mathbf S$ for all of the calculations.  The full numerical calculations reveal that mass state hopping is more adiabatic than our zeroth order approximation would lead us to believe, differing by $\approx 5-10\,\%$ from the hopping probability associated with the evolution of $\mathbf S^{(0)}$.

\begin{figure}
\centering
\includegraphics[scale=0.72]{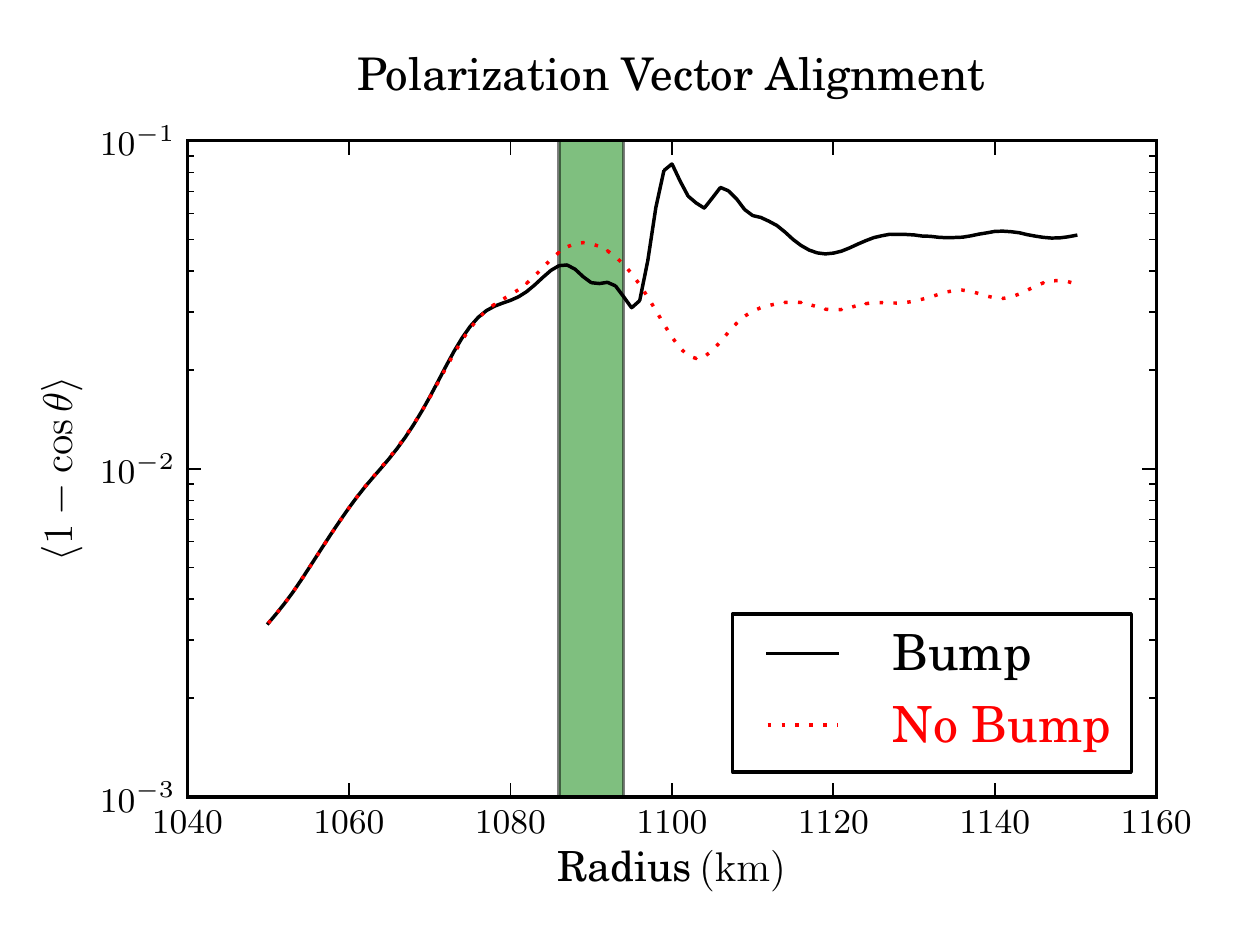}
\caption{The average alignment of individual neutrino polarization vectors, $\mathbf s_\omega$, with the collective polarization vector, $\mathbf S$.  The solid black line shows the \lq\lq Bump\rq\rq\ profile from Refs.~\cite{Nomoto84,Nomoto87} and the dotted red line shows the modified \lq\lq No Bump\rq\rq\ profile.  The shaded region indicates the physical position of the Helium burning shell density feature present in the Bump profile.}
\label{fig:Avrad}
\end{figure}

The reason that the full calculations exhibit less mass state hopping can be found in the imperfect alignment of individual $\mathbf s_\omega$ with $\mathbf S$.  The Neutrino Enhanced MSW model predicts that individual $\mathbf s_\omega$ tend to stay aligned with $\mathbf S$ as the collective mode passes through resonance and subsequently begin to orbit around $\mathbf H_{\rm v}$ as the system transitions to the regular precession mode.  In Figure~\ref{fig:Avrad} we show the average opening angle between individual $\mathbf s_\omega$ and $\mathbf S$, taken to be $\cos{\theta} = \left(\hat{\mathbf s}_\omega \cdot \hat{\mathbf S} \right)$, for the Bump and No Bump profiles.  On average individual $\mathbf s_\omega$ and $\mathbf S$ remain aligned to within a few percent throughout the resonance region in both the Bump and No Bump simulations, which shows that this is indeed the correct physical picture.  However, as we have mentioned previously, the individual $\mathbf s_\omega$ pass through resonance at slightly higher densities than $\mathbf S^{(0)}$.  The individual $\mathbf s_\omega$ are slightly misaligned and this means that when the individual neutrino states are at resonance, i.e. $\cos{2\theta_\omega}=0$, the collective mean field $\mathbf S$ is not yet at resonance itself, i.e. $\cos{2\theta_{\rm m}}\neq 0$. 

From Equation~\ref{costhw} we see that for an individual neutrino state at resonance the adiabaticity of the mass state hopping will not be determined entirely by the matter potential if $\cos{2\theta_{\rm m}}\neq 0$, which is precisely the result we recover from Equation~\ref{BDSres} when $A>B$ with $B\neq0$.  (By contrast, the flavor evolution through resonance for $\mathbf S^{(0)}$ is determined entirely by the matter potential.)  Individual $\mathbf s_\omega$ experience some fraction of the neutrino self-coupling potential $\mu$.  This comes from the fact that our approximation $g\left(\omega\right) \approx \delta\left(\omega - \langle\omega\rangle\right)$ is a rather gross approximation.  In reality, the function $g\left(\omega\right)$ has a finite width.  However, we find that this approximation produces results that match well with our calculations and individual $\mathbf s_{\omega}$ track the evolution of $\mathbf S^{(0)}$ more closely than one might expect given the width of our initial $\nu_{\rm e}$ distribution.

A serious source of potential error in interpreting the swap signal for the neutronization neutrino burst comes from rapid fluctuations in the matter potential.  We have shown the evolution of the system follows $\mathbf S^{(0)}$ closely.  In turn, $\mathbf S^{(0)}$ experiences only the ordinary MSW effect in it's flavor evolution.  This is illustrated clearly by the similarities in the observed mass state 3 survival probabilities and the trajectories of $\mathbf S$ and $\mathbf S^{(0)}$ in Figures~\ref{fig:ALGNBUMP} and ~\ref{fig:ALGNNOBUMP}.  For the No Bump profile $P^{\rm Observed}_{\rm H} = 0.76$, and $P^{\mathbf S^{(0)}}_{\rm H} =  0.68$.  For the Bump profile $P^{\rm Observed}_{\rm H} = 0.85$, and $P^{\mathbf S^{(0)}}_{\rm H} =  0.80$.  However, as we have mentioned in the previous section, this result for the Bump profile is in stark disagreement with the prediction of the double exponential Landau-Zener hopping probability of Equation~\ref{Phop}.  

As has been know for some time, turbulent matter density fluctuations can produce flavor \lq\lq depolarization\rq\rq\ for MSW neutrino flavor transformation~\cite{Sawyer:1990lr,Loreti:1994qy,Loreti:1995fk,Balantekin:1996uq,Friedland:2006lr,Kneller:2008rt,Gava:2009yq,Kneller:2010vn,Kneller:2010ys}.  Broadly, these turbulent fluctuations can produce mass state hopping that does not agree with what one would deduce from Eqn.~\ref{Phop} using the gross scale height of the matter potential.  

 \begin{figure}
\centering
\includegraphics[scale=0.72]{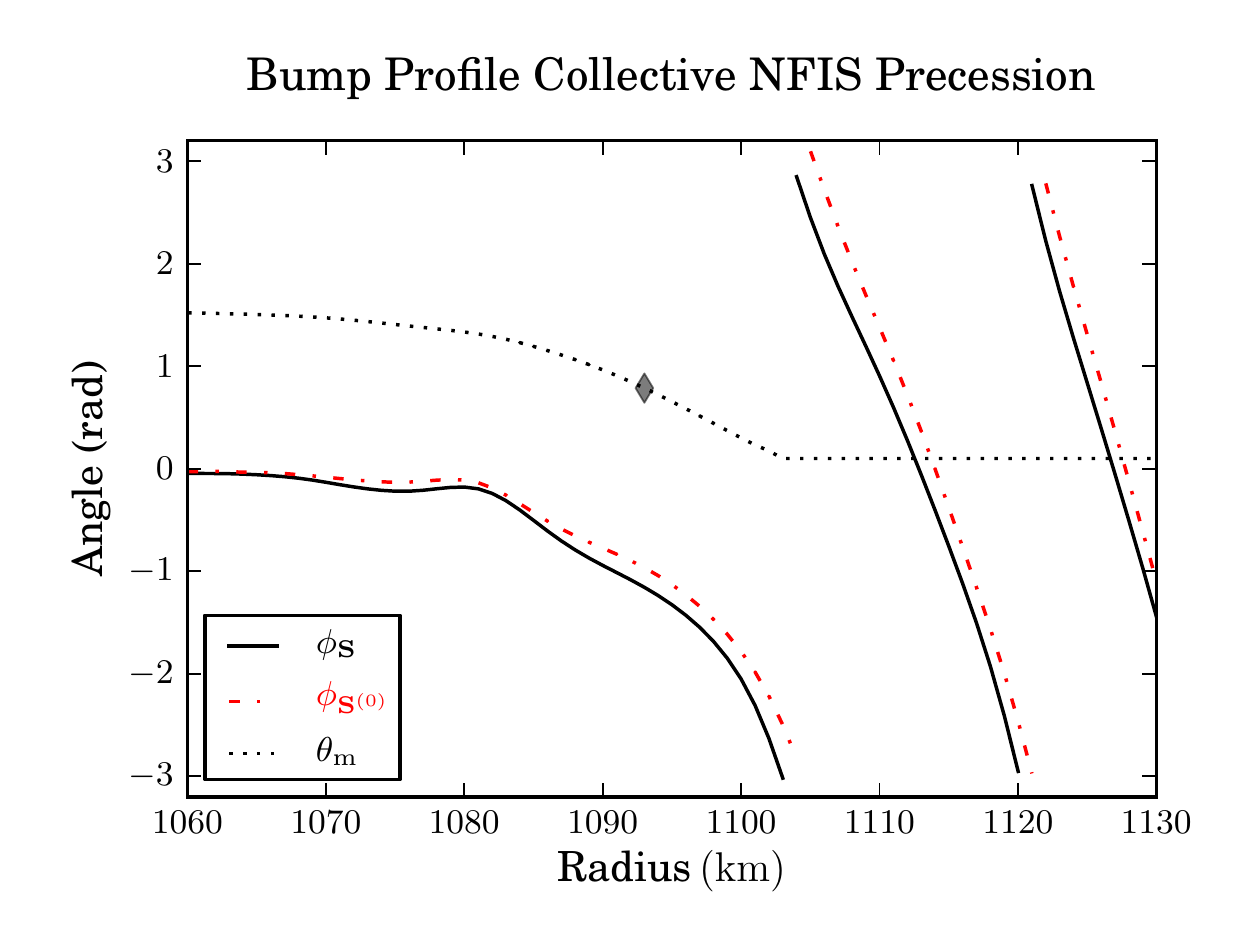}
\caption{No Bump:  The precession of the collective polarization vector, $\mathbf S$, and the zeroth order effective field $\mathbf S^{(0)}$, about the $\hat{\mathbf e}^{\rm f}_{\rm z}$ axis as neutrinos move through resonance in the No Bump profile.  The solid line shows the azimuthal angle, $\phi_{\mathbf S}$, for the collective polarization vector $\mathbf S$.  The dashed-dotted line shows the azimuthal angle, $\phi_{\mathbf S^{(0)}}$, for the zeroth order effective field $\mathbf S^{(0)}$.  The dotted line shows the value of $\theta_{\rm m}$ for $\mathbf S$ and $\mathbf S^{(0)}$ for reference, with a diamond symbol located at $\theta_{\rm m} = \pi /4$ where the system is at resonance.}
\label{fig:PhiNoBump}
\end{figure}
 
  \begin{figure}
\centering
\includegraphics[scale=0.72]{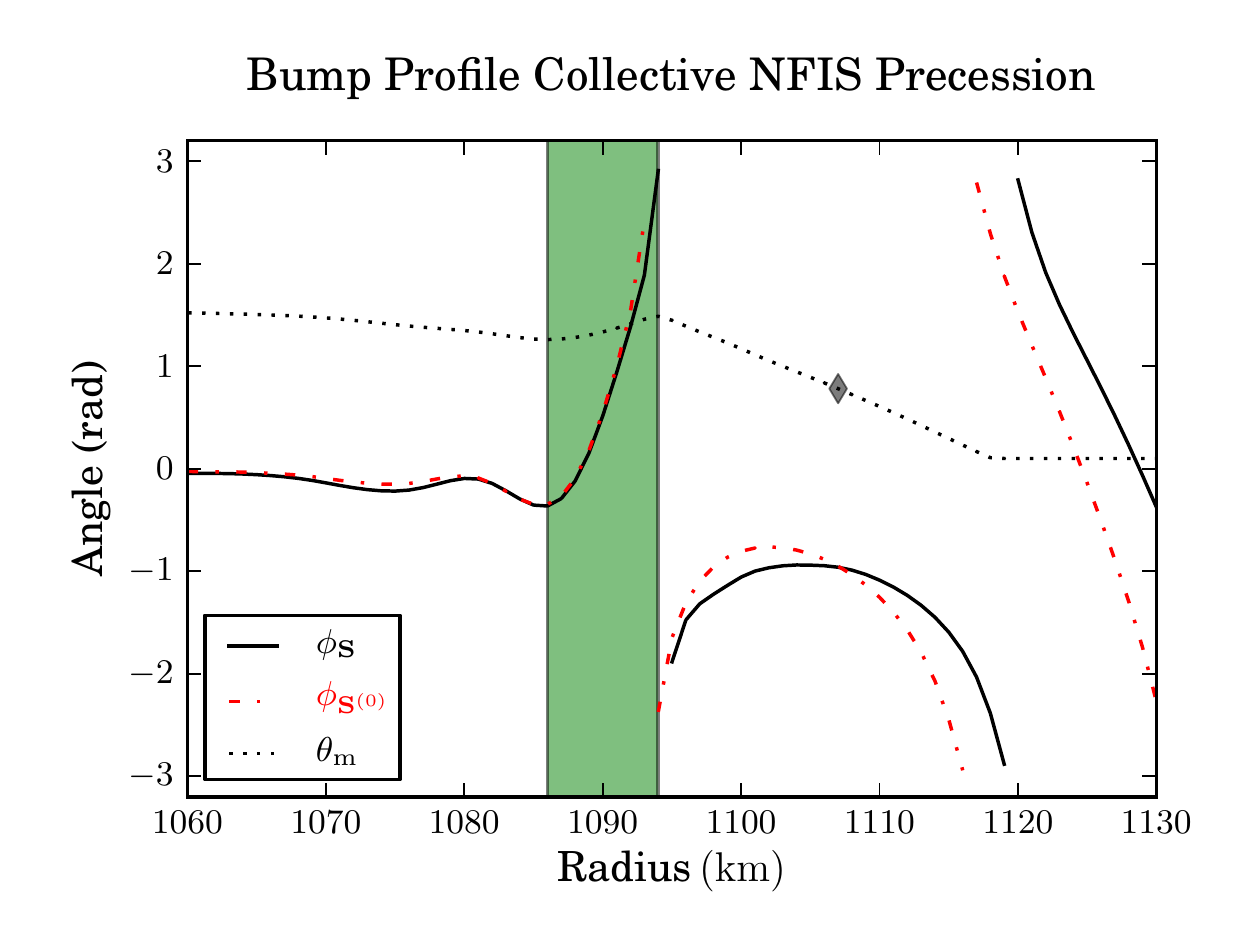}
\caption{Bump:  The precession of the collective polarization vector, $\mathbf S$, and the zeroth order effective field $\mathbf S^{(0)}$, about the $\hat{\mathbf e}^{\rm f}_{\rm z}$ axis as neutrinos move through resonance in the Bump profile.  The solid line shows the azimuthal angle, $\phi_{\mathbf S}$, for the collective polarization vector $\mathbf S$.  The dashed-dotted line shows the azimuthal angle, $\phi_{\mathbf S^{(0)}}$, for the zeroth order effective field $\mathbf S^{(0)}$.  The dotted line shows the value of $\theta_{\rm m}$ for $\mathbf S$ and $\mathbf S^{(0)}$ for reference, with a diamond symbol located at $\theta_{\rm m} = \pi /4$ where the system is at resonance.  The shaded region indicates the physical position of the Helium burning shell density feature present in the Bump profile.}
\label{fig:PhiBump}
\end{figure}

By considering the derivative of Eqn.~\ref{S_zero_EOM} it can be seen that there is a restoring force that causes $\mathbf S^{(0)}$ to orbit $\mathbf H_{\rm MSW}$ as the system evolves.   Another term in the derivative of Eqn.~\ref{S_zero_EOM} drives precession of $\mathbf S^{(0)}$ about the $\hat{\mathbf e}^{\rm f}_{\rm z}$ axis any time the matter potential is changing non-adiabatically.  
 
 In the traditional MSW framework, this precession is dependent on the energy of each neutrino, as the neutrino energy sets the relative size of the terms in Eqn.~\ref{S_zero_EOM}'s derivative.  As a result, if the matter density profile is turbulent, the alignments of individual neutrino NFIS's can be scattered throughout the flavor space, hence the term \lq\lq depolarization\rq\rq .  
 
The matter driven precession figures into the hopping probability given by Eqn.~\ref{Phop} because the double exponential hopping probability is derived from the Landau-Zener two level hopping problem using the boundary condition that the precession of $\mathbf S^{(0)}$ originating from this term  can be taken to be zero before the system approaches resonance~\cite{Haxton:1986uq,Haxton:1987yq}.  If there is significant matter driven precession for $\mathbf S$ and $\mathbf S^{(0)}$ prior to resonance, Eqn.~\ref{Phop}  will not be the appropriate analytic solution for the Landau-Zener hopping probability.
 
These issues can be explored by examining the projection of the collective field $\mathbf S$ (or $\mathbf S^{(0)}$) in the $\hat{\mathbf e}^{\rm f}_{\rm x}-\hat{\mathbf e}^{\rm f}_{\rm y}$ plane.  Define the angle made by this projection and $\hat{\mathbf e}^{\rm f}_{\rm x}$ to be $\phi$.  Figure~\ref{fig:PhiNoBump} shows the value of $\phi$ in the No Bump profile for $\mathbf S$ and $\mathbf S^{(0)}$ through the resonance region.  Prior to resonance, when $\theta_{\rm m} \backsimeq \pi/2$, there is no appreciable precession of either $\mathbf S$ or $\mathbf S^{(0)}$.  This is not particularly surprising, as the observed hopping probability of $\mathbf S^{(0)}$ matches exactly with Eqn.~\ref{Phop}, with $\mathbf S$ exhibiting slightly less hopping as we have discussed above.
 
 Figure~\ref{fig:PhiBump}, where we show the evolution of $\phi$ for $\mathbf S$ and $\mathbf S^{(0)}$ in the Bump profile, exhibits very different phenomenology.  The system starts out identically to the configuration of the neutrinos in the No Bump profile, with no matter driven precession before the bump is reached.  However, the sudden increase in the local matter density at $r \sim 1086\, \rm km$ brought on by the helium burning shell drives rapid precession of $\mathbf S$ and $\mathbf S^{(0)}$ in a counter clock-wise direction.  At the same time, the value of $\theta_{\rm m}$ is brought back up to the pre-resonance value of $\theta_{\rm m}\backsimeq \pi/2$.  This resets the clock, so to speak, on $\mathbf H_{\rm MSW}$ and the boundary conditions for $\mathbf S$ and $\mathbf S^{(0)}$.  The collective NFIS is clearly precessing rapidly at the boundary of the MSW resonance region, meaning that our use of the double exponential hopping formula is not appropriate.
 
 This phenomenon is a simplistic example of the turbulence driven flavor depolarization discussed in Ref.~\cite{Loreti:1995fk,Friedland:2006lr,Kneller:2010vn,Kneller:2010ys}, where we are following the evolution of a single NFIS, $\mathbf S^{(0)}$, as it is scattered to a random position in flavor space.  What is surprising about our calculation is that we have observed that the collective Neutrino Enhanced MSW effect persists and maintains coherence in multi-angle calculations, in spite of rough handling by the matter potential.  The collective NFIS, $\mathbf S$ (and consequently the individual neutrino $\mathbf s_{\omega}$'s), tracks the evolution of $\mathbf S^{(0)}$ and achieves a final alignment in flavor space that is not relatable to the simplistic prediction of Eqn.~\ref{Phop}.  Our calculations suggest that the Neutrino Enhanced MSW effect is susceptible to the same turbulence driven precession as the ordinary MSW effect.  
 
This is unfortunate from the perspective of attempts to use the swap signal to probe the matter density profile.  In the particular case of the neutronization neutrino pulse, the resultant swap energy may or may not be relatable to the matter density profile, depending on the \lq\lq smoothness\rq\rq\ of the actual matter density profile of the collapsing star.  In the case of the Bump profile, our observed survival probability of $P^{\rm Bump}_{\rm H} = 0.852$ would lead an observer to infer a matter density scale height of $R^{\rm Bump,\ \rm inferred}_{\rm H} = 45.9\,\rm km$, compared to the actual scale height of $R^{\rm Bump}_{\rm H} = 25.3\,{\rm km}$, an $85\%$ error.  In principle, this fluctuation driven precession might alter the survival probability $P_{\rm H}$ and, consequently, push the swap energy up or down depending on the particular realization of the density fluctuations~\cite{Loreti:1995fk,Friedland:2006lr,Kneller:2010vn,Kneller:2010ys}.

In the example we study here, the low energy neutrinos experiencing multiple MSW resonances would be almost invisible to neutrino detectors designed to collect supernova neutrinos, since these detectors might have low energy thresholds of $\sim 5 - 10\,\rm MeV$.  However these neutrinos provide our only clue about matter density fluctuations.  While we have been able to use our detailed knowledge of the neutrino flavor states as they evolved through the resonance region to correctly deduce what has transpired, an observer on the Earth would not have access to such privileged information.
 
\section{Conclusion}
We have elucidated the rich interplay of collective neutrino oscillations and the underlying matter density profile in a supernova.  Our calculations reveal a heretofore unrecognized aspect of collective oscillations, namely the matter density fluctuation-driven precession of the collective ensemble of neutrino NFIS's (e.g. please view the movies located at~\cite{Cherry:web}).  The implications of this for neutrino-detection astrophysics are somewhat negative.

In particular, we have shown that during the neutronization neutrino pulse epoch of an O-Ne-Mg core-collapse supernova there are strong limitations on an observer's ability to use the swap signal to probe fluctuations in the matter density profile of the collapsing star (e.g. the density ledges left by fossil burning shells).  If the matter density profile of the progenitor O-Ne-Mg star is not smooth, reverse engineering the swap energy to find a matter density scale height may not necessarily be possible.  

The problem hinges on an observer's ability to detect neutrinos which have energies low enough that their flavor evolution histories would be disconnected from collective flavor oscillations in the presence of matter density fluctuations.  If low energy features such as the one seen in Figure~\ref{fig:Pbump} are observed during the neutronization neutrino pulse epoch, an observer could at the very least deduce the presence of a fluctuating matter density profile.  However, this requires either a neutronization neutrino pulse that has average neutrino oscillation frequency, $\langle \omega \rangle$, larger than expected from progenitor models, or neutrino detectors that have neutrino energy detection thresholds below $5\, {\rm MeV}$.

The model presented here suggests further investigation.  It is likely that the Neutrino Enhanced MSW effect could provide a secondary probe of the neutrino luminosity during the neutronization burst.  Furthermore, $\nu_{\rm e}$ fluxes during this epoch with $\langle \omega \rangle$ different from the value that we have used here may also produce unique flavor transformation signatures that could potentially be used as an independent measure of the $\nu_{\rm e}$ temperature. 

It should be noted that these effects are limited to supernova neutrino systems that experience the Neutrino Enhanced MSW effect, which is not universal to all supernova neutrino systems.  As the neutrino flux from the proto-neutron star evolves past the neutronization neutrino burst phase, entirely different collective effects are capable of producing swaps which may not be susceptible to turbulent matter density fluctuations.  

Nevertheless, the study presented here gives new insights into collective neutrino oscillations in supernovae.  This and other neutrino flavor oscillation studies may help drive synergy between the laboratory neutrino physics enterprises and observational and theoretical astrophysics.

\section{Acknowledgments}
This work was supported in part by NSF grant PHY-06-53626 at UCSD, DOE grant DE-FG02-87ER40328 at the UMN, and by the DOE Office of Nuclear Physics, the LDRD Program and Open Supercomputing at LANL, and an Institute of Geophysics and Planetary Physics/LANL minigrant.  We would like to thank the topical collaboration for neutrino and nucleosynthesis in hot and dense matter at LANL and the New Mexico Consortium.

\bibliography{allref}

\end{document}